\documentclass[10 pt]{article}
\usepackage{amsmath,amsfonts,amssymb,bm}
\usepackage{float}
\usepackage{subfigure}
\usepackage{graphicx}
\usepackage{wrapfig}
\usepackage[margin=1in]{geometry}

\begin{document}
\title{On the Preservation of the Local Properties of Time in General Relativity}
\author{Robert D. Bock\footnote{rdbock@gmail.com}}
\date{\today}

\maketitle

\begin{abstract}
\noindent 
We present a theory of gravity based on Einstein's general relativity that is motivated by the paradoxes associated with time in relativistic rotating frames and certain exact solutions of Einstein's equations.  We show that we can resolve these paradoxes with a single postulate, namely, the preservation of the local properties of time.  This postulate forces the introduction of four additional degrees of freedom into the space-time structure whose transformation properties are defined to preserve the local properties of time under arbitrary four-dimensional coordinate transformations.  We identify these additional degrees of freedom with the electromagnetic potentials, which thereby acquire a space-time interpretation.  The resulting theory is similar to the traditional Einstein-Maxwell theory, differing primarily in the emergence of a coupling between the gravitational and electromagnetic fields.  We show that the proposed reformulation of general relativity provides a rich framework for addressing a number of outstanding problems in contemporary gravitational physics.


\end{abstract}

                             
\section{\label{sec:introduction}Introduction}
Einstein's theory of general relativity \cite{einstein1916} has enjoyed overwhelming success since its inception, predicting phenomena that have been observed in our solar system and beyond, including the perihelion precession of Mercury, relativistic effects in the Hulse-Taylor binary pulsar B1913+16, and the existence of black holes.  Despite its successes, a number of outstanding issues persist, both in the special and general theories.  These include\footnote{This list is by no means exhaustive.}: the paradoxes associated with relativistic rotating frames of reference \cite{relrotbook}; the existence of closed timelike curves \cite{lobocrawford} in numerous exact solutions of the Einstein field equations; the inability to unify the electromagnetic and gravitational fields \cite{einstein1945,pais1982}; the absence of a consistent quantum theory of gravity \cite{woodard2009}; the need to postulate dark matter in order to resolve the observed acceleration discrepancies in astrophysical systems \cite{turner,silk,brainerdetal}; the dark energy problem in the $\Lambda\mathrm{CDM}$ model \cite{copelandetal}; and, the apparent anomalous acceleration observed in radio Doppler and ranging data from the Pioneer missions \cite{andersonetal1988}. 

Consequently, many theories have been proposed that attempt to reconcile relativity theory with these outstanding problems. The earliest examples of such theories include the unified field theories of Einstein \cite{einstein1945,pais1982} and others \cite{kaluza1921,weyl1922,klein1926,eddington1954,schrodinger1950,pauli1958}.  More recent examples include new formulations of special relativity in rotating reference frames \cite{relrotbook}.  In gravitational physics, contemporary examples include modifications of general relativity that may eliminate the need for dark matter \cite{finzi,tohline,sanders1984,sanders1986,goldman,kuhnandkruglyak,milgrom1983a,milgrom1983b,milgrom1983c,bekenstein2004} and dark energy (see Reference \cite{copelandetal} for a review of the various theories that incorporate general forms of dark energy such as quintessence, K-essence, tachyon, phantom, and dilatonic models).  To date, none of these theories have superseded Einstein's original theory \cite{einstein1916}, in either the special or general formulations.

In the following we propose a new formulation of the gravitational field equations based on general relativity that addresses a number of these aforementioned issues.  It is based on the preservation of the local properties of time under arbitrary four-dimensional coordinate transformations.  We motivate the theory by observing that the well-known paradoxes associated with time in relativistic rotating frames \cite{relrotbook} and certain exact solutions of Einstein's equations \cite{lobocrawford,nahin} are resolved by demanding that the terms associated with physical space and time measurements remain separately invariant under the full set of coordinate transformations, and not just the restricted set of the traditional theory.  As a result, we are forced to introduce four new degrees of freedom into the space-time structure, which are identified with the electromagnetic field and whose contribution to the space-time interval sums identically to zero.  The resulting field equations resemble the traditional Einstein-Maxwell theory, however, a coupling between the gravitational and electromagnetic fields emerges.  We show that this theory provides a rich framework for addressing a number of outstanding issues in contemporary gravitational physics.

\section{\label{sec:paradoxes}Paradoxes Associated with Time in Relativity}
In this section we briefly discuss two closely-related paradoxes associated with time in Einstein's theory of relativity: the time discontinuity paradox in relativistic rotating frames \cite{relrotbook} and the existence of Closed Timelike Curves (CTCs) \cite{lobocrawford,nahin} in certain exact rotating solutions of Einstein's equations.  

\subsection{Time Discontinuity Paradox}
The time discontinuity (or time lag) arises when one tries to establish standard simultaneity along a closed curve in a rotating coordinate system.  Upon traversing a complete circuit in such a frame of reference an observer discovers that a clock situated at the curve's orgin is not synchronized with itself.  This is often treated in the context of special relativity alone.  According to the traditional viewpoint (see, for example, Refs. \cite{dieks, weber,anandan,bergiaguidone,cranoretal,rizzitartaglia}) special relativity is valid in rotating frames of reference and the time discontinuity is only an apparent problem. This traditional approach maintains that multiple clock readings at a given event, depending on the chosen synchronization procedure, are indeed acceptable.  Furthermore, it is argued that the time gap is no more problematic than the discontinuity in time at the International Date Line or the coordinate discontinuity in angle at 2$\pi$.  On the other hand, many authors have questioned the validity of special relativity in rotating frames of reference and have attempted to modify Einstein's postulates for rotational motion.  For example, Klauber \cite{klauber2} and independently, Selleri \cite{selleri2} contend that the synchronization procedure cannot be chosen freely for the rotating frame and propose a unique (non-Einstein) synchronization along the circumference.  

Consider a Minkowski space-time with cylindrical coordinates $\{T,R,\Phi,Z\}$.  The line element is given by:
\begin{equation}
\label{flatlineelement}
ds^2=c^2dT^2-dR^2-R^2d\Phi^{2}-dZ^2.
\end{equation}
The coordinate transformation from the laboratory frame $\{T,R,\Phi,Z\}$ to the rotating frame $\{t,r,\phi,z\}$ is given by:
\begin{equation}
\label{coordtransformation}
T=t, R=r, \Phi=\phi+\omega t, Z=z, 
\end{equation}  
where $\omega$ is the angular velocity of the rotating system as observed from the laboratory frame.  Substituting (\ref{coordtransformation}) into (\ref{flatlineelement}) gives:
\begin{equation}
\label{rotlineelement}
ds^2=\gamma^{-2}c^{2}dt^{2} - 2c\beta rd\phi dt-dr^{2}-r^{2}d\phi^{2} -dz^{2},
\end{equation}
where $\beta=\omega r /c<1$ and $\gamma = (1-\beta^2)^{-1/2}$. Note that the condition $\beta<1$ is arbitrarily imposed on the coordinate transformation.

A self-consistent problem emerges when one attempts to define simultaneity globally in the rotating frame of reference.  Consider two clocks, A and B, separated by the infinitesimal distance $rd\phi$ along the circumference in the rotating frame.  In order to define standard simultaneity between the two (infinitesimally near) clocks the time on clock B must be adjusted by the amount \cite{landaulifshitz}:
\begin{equation}
\label{localtimelag}
c\Delta t = -\beta\gamma^{2}rd\phi.
\end{equation}  
The well-known expression for the time discontinuity is obtained by integrating around the entire circumference of the circle:
\begin{equation}
\label{timegap}
\Delta t = -\frac{2\pi\beta\gamma^{2}r}{c}.
\end{equation}
Thus, if one sends a light ray from a clock A around the entire circumference of the circle, establishing standard simultaneity along the way, then one discovers that the clock at A is not synchronized with itself.  While nearby clocks on an open curve can be synchronized by adjusting the readings of the various clocks according to Eq. (\ref{localtimelag}), this procedure cannot be extended globally since $\Delta t$ in Eq. (\ref{localtimelag}) is not a total differential in $r$ and $\phi$.  That is to say, the synchronization procedure is path dependent in the rotating frame of reference. 

\subsection{Closed Timelike Curves (CTCs)}
Closed timelike curves are also the subject of much debate in Einstein's theory of relativity.  A CTC is a future directed timelike curve in the space-time manifold that runs smoothly back into itself.  As is well known \cite{nahin}, the existence of CTCs suggests that time travel is compatible with general relativity since an observer may evolve in time within the future light cone and return to an event that coincides with an earlier departure from that event.  A number of exact solutions of the Einstein field equations exhibit nontrivial CTCs, including a rapidly rotating infinite cylinder \cite{vanstockum, tipler}, the G\"{o}del universe \cite{godel}, a Kerr black hole \cite{carter}, and spinning cosmic strings \cite{deseretal,gott}.  While the G\"{o}del universe, the cosmic strings and the van Stockum cylinder all possess properties that may be deemed unphysical, the low angular momentum Kerr black hole is believed to possess physical relevance - it is the unique final state of gravitational collapse \cite{wald}.  Therefore, CTCs cannot be dismissed simply as mathematical curiosities.  Furthermore, the proliferation of new solutions that exhibit CTCs \cite{mallett,bonnor2,bonnor3,bonnorward,bicakpravda} suggests that their appearance in general relativity poses a critical problem to the foundations of physics \cite{bonnor}.

Hawking \cite{Hawking} has suggested that quantum effects prevent the emergence of CTCs.  In particular, he showed that divergences in the energy momentum tensor caused by vacuum polarization effects create singularities prior to the appearance of CTCs.  Based on these results Hawking proposed the chronology protection conjecture:  the laws of physics do not allow the appearance of CTCs.  Kim and Thorne \cite{kimthorne} have suggested otherwise, namely, that the divergences in the energy momentum tensor may be cut off by quantum gravitational effects.  Without a well-defined theory of quantum gravity this matter is still open to debate \cite{lobocrawford}.

\section{\label{sec:resolution}Resolution of the Paradoxes}
Neither the time discontinuity nor CTCs have been experimentally observed.  According to the traditional viewpoint the time discontinuity paradox is only an apparent problem, whereas the existence of CTCs is more fundamental.  Deviating from this viewpoint, we assert that both of these paradoxes are real and are evidence of a fundamental crisis in relativity theory.  Therefore, the inability to experimentally realize both the time discontinuity and CTCs forces us to modify general relativity in a manner that is consistent with our physical experience of time.  In this section we show that the modification we seek follows from a single postulate, namely, the preservation of the local properties of time.  In addition, we show that four new degrees of freedom are required to satisfy this postulate.

The space-time interval of conventional general relativity is given by\footnote{In this paper, Greek indices will run from $0\ldots 3$, lower-case Latin indices will run from $1\ldots 3$, and $c=1$, unless otherwise stated.}:
\begin{equation}
\label{eq:metric1}
ds^2 = g_{\mu\nu}dx^{\mu}dx^{\nu},
\end{equation}
where $g_{\mu\nu}$ is symmetric and $ds^2$ is invariant under the full set of general coordinate transformations:
\begin{eqnarray}
\label{eq:general_transformations}
x^{\prime 0} &=& f\left(x^{\alpha}\right) \nonumber \\
x^{\prime i} &=& x^{\prime i}\left(x^{\alpha}\right).
\end{eqnarray}
As is well-known \cite{landaulifshitz,moller1955}, the space-time interval may be decomposed into two separate terms representing the contributions of physical time and space measurements, respectively:
\begin{equation}
\label{eq:metric2}
ds^2 = d\sigma^2 - dl^2,
\end{equation}
where
\begin{eqnarray}
\label{eq:physical_time_space}
d\sigma^2 &\equiv& g_\mu g_\nu dx^\mu dx^\nu \nonumber \\
-dl^2   &\equiv&\left(g_{\mu\nu} - g_\mu g_\nu\right)dx^\mu dx^\nu
\end{eqnarray}
and we have introduced the notation:
\begin{equation}
g_u \equiv \frac{g_{0\mu}}{\sqrt{g_{00}}}.
\end{equation}
Note that $-dl^2$ is often written as a sum over the spatial coordinates only:
\begin{equation}
-dl^2   \equiv\left(g_{ij} - g_i g_j\right)dx^i dx^j.
\end{equation}
However, the definition in Equation (\ref{eq:physical_time_space}) is equivalent since $g_{00}-g_0g_0 = g_{0i}-g_0g_i=0$.  Note that by writing the spatial distance as in Equation (\ref{eq:physical_time_space}), the relationship (\ref{eq:metric2}) becomes more explicit in Equation (\ref{eq:physical_time_space}):
\begin{equation}
\label{eq:physical_space}
-dl^2 = ds^2 - d\sigma^2.
\end{equation}
It is important to emphasize that Equation (\ref{eq:metric2}) follows from Equation (\ref{eq:metric1}) by adding and subtracting the quantity $g_i g_jdx^i dx^j$ (adding zero):
\begin{eqnarray}
\label{eq:addzero}
ds^2 &\rightarrow& \left[g_{00}\left(dx^{0}\right)^2 + 2g_{0i}dx^0dx^i+ g_i g_jdx^i dx^j\right] +\left[g_{ij}dx^i dx^j - g_i g_jdx^i dx^j \right] \nonumber \\
&=&ds^2 + 0.
\end{eqnarray}

The terms $d\sigma^2$ and $-dl^2$ are not separately invariant under the full set of general coordinate transformations (\ref{eq:general_transformations}), but are each invariant under a `gauge transformation of the gravitational potentials' \cite{moller1955}, the restricted group of transformations for which the functions defining the transformations of the spatial coordinates do not depend on time:
\begin{eqnarray}
\label{eq:restricted_transformations}
x^{\prime 0} &=& f\left(x^{\alpha}\right) \nonumber \\ 
x^{\prime i} &=& x^{\prime i}\left(x^{j}\right).
\end{eqnarray} 
This is the reason the time discontinuity paradox emerges.  The term that is responsible for the time discontinuity (\ref{localtimelag}) is a result of the time dependence of the angular coordinate transformation in Equation (\ref{coordtransformation}).  In other words, the local properties of time are not preserved under the transformation (\ref{coordtransformation}) from the laboratory frame to the rotating frame.  Similarly, for `permanent' gravitational fields, the local properties of time are not fixed for all matter distributions, and therefore CTCs can emerge for valid solutions of Einstein's equations even when they are not observed in standard systems of reference.  There is no principle nor mechanism in gravitational theory proper that preserves the local properties of time, given a standard system of reference.

\textit{Therefore, we advance the postulate that the local properties of time are externally constrained in general relativity and must be preserved in the formulation of the theory.}  According to this postulate, the local properties of time must remain invariant under the general set of coordinate transformations (\ref{eq:general_transformations}) and not just the restricted set (\ref{eq:restricted_transformations}) of traditional general relativity.  To be sure, this postulate does not preclude a physical time with local properties different from that observed in our standard system of reference (for example, it does not preclude path-dependent synchronization).  Rather, the postulate forbids a general coordinate transformation (\ref{eq:general_transformations}) from modifying the local properties of time, whatever they may be prior to the transformation.  This will be our primary point of departure from traditional relativity, which, as we have seen above possesses a non-invariant $d\sigma$.  In the remainder of the paper, we reformulate general relativity to satisfy this postulate and examine its consequences.

The term $d\sigma = g_\mu dx^\mu$ is not a scalar invariant because $g_\mu$ does not transform as a four-vector.  However, it can be made an invariant if we make the following substitution in Equation (\ref{eq:physical_time_space}):
\begin{equation}
\label{eq:psi_definition}
g_\mu \rightarrow \psi_\mu \equiv g_\mu + \phi_\mu,
\end{equation}
where $\phi_\mu$ are four new degrees of freedom whose transformation properties are defined so that $\psi_\mu$ transforms as a four-vector.  This substitution yields:
\begin{eqnarray}
\label{eq:physical_time_space2}
d\sigma^2 &=& \psi_\mu \psi_\nu dx^\mu dx^\nu \nonumber \\
-dl^2   &=&\left(g_{\mu\nu} - \psi_\mu \psi_\nu\right)dx^\mu dx^\nu.
\end{eqnarray}
Since $ds^2 = d\sigma^2 - dl^2$, the new terms that result from the introduction of $\phi_\mu$ automatically subtract out so that the net effect on the space-time interval results in the addition of zero (cf.\ Equation (\ref{eq:addzero})):
\begin{eqnarray}
\label{eq:new_line_element}
ds^2 &=& g_{\mu\nu} dx^\mu dx^\nu \nonumber \\
     &=&g_\mu g_\nu dx^{\mu}dx^{\nu} +\left(g_{\mu\nu} - g_\mu g_\nu\right)dx^{\mu} dx^{\nu}\nonumber \\ 
     &=& \left[\left(g_\mu + \phi_\mu \right)\left(g_\nu + \phi_\nu \right)dx^{\mu}dx^{\nu}\right] 
     +\left[ g_{\mu\nu} - \left(g_\mu + \phi_\mu \right)\left(g_\nu + \phi_\nu \right) \right]dx^{\mu} dx^{\nu} \nonumber \\
     &=& g_{\mu\nu} dx^\mu dx^\nu + 0.
\end{eqnarray}

The field $\phi_\mu$ can be viewed as a `gauge'\footnote{We use the term `gauge' loosely since the introduction of the field does not involve a derivative.} field whose function is to preserve the local properties of time in a relativistic system of reference.  The space-time interpretation of the field $\phi_\mu$ follows by examining the definition of simultaneity in general relativity \cite{landaulifshitz}.  Following Reference \cite{landaulifshitz} we consider the propagation of a signal from point $B$ in space (with coordinates $x^i + dx^i$) to a point $A$ infinitely near (with coordinates $x^i$) and then returning back to $B$ over the same path.  The time of arrival at $A$ is $x^0$ and the times of departure and arrival at $B$ are $x^0+dx_0^{(1)}$ and  $x^0+dx_0^{(2)}$, respectively, where $dx_0^{(1)}$ and $dx_0^{(2)}$ are the two roots for $dx^0$ of the equation $ds^2=0$:
\begin{equation}
\label{eq:dx0_1_2}
dx_0^{(1),(2)} = -\frac{1}{g_{00}}\left[g_{0i}dx^i \mp \sqrt{\left(g_{0i}g_{0j}-g_{ij}g_{00} \right)dx^idx^j} \right].
\end{equation}
The quantities $dx_0^{(1),(2)}$ correspond to the coordinate time intervals for propagation in the two directions between $A$ and $B$.  Landau and Lifshitz \cite{landaulifshitz} point out that the time at $B$ which is simultaneous with the the time of arrival of the light signal at $A$, $x_0$, is shifted in coordinate time by the amount: 
\begin{equation}
\label{eq:old_simultaneity}
d x^0 = \frac{1}{2}\left(dx_0^{(2)} + dx_0^{(1)}\right).
\end{equation}
Converting this to a proper time interval and substituting Equation (\ref{eq:dx0_1_2}) this becomes:
\begin{equation}
\label{eq:proper_time_difference}
\Delta \tau = \sqrt{g_{00}}dx^0=g_0dx^0 = -g_i dx^i.
\end{equation}
Therefore, in traditional general relativity local simultaneity is defined by the vanishing of the quantity $g_\mu dx^\mu $:
\begin{equation}
g_\mu dx^\mu = 0.
\end{equation}
Note that this definition of local simultaneity is not generally covariant.  However, a generally covariant definition of local simultaneity can be obtained by redefining the simultaneity condition so that the proper time difference in simultaneous events at $A$ and $B$ is a function of the field $\phi_\mu$ according to:
\begin{equation}
\label{eq:new_simultaneity}
\sqrt{g_{00}}dx^0 = \frac{\sqrt{g_{00}}}{2}\left(dx_0^{(2)} + dx_0^{(1)}\right) - \phi_\mu dx^\mu,
\end{equation}
where $\phi_\mu$ transform as defined above.  As a result, local simultaneity is defined by the covariant relationship:
\begin{equation}
g_\mu dx^\mu = -\phi_\mu dx^\mu.
\end{equation}
This equation is equivalent to that obtained by setting $d\sigma=0$ in Equation (\ref{eq:physical_time_space2}).  Hence, the field $\phi_\mu$ enables local simultaneity to be defined covariantly throughout space-time.  Traditional general relativity defines local simultaneity via Equation (\ref{eq:old_simultaneity}), and therefore does not preserve the local properties of time under general coordinate transformations.  We see that by replacing Equation (\ref{eq:old_simultaneity}) with Equation (\ref{eq:new_simultaneity}) we can satisfy the postulate of the preservation of the local properties of time. 

We can rewrite Equation (\ref{eq:new_simultaneity}) in order to reveal explicitly the space-time roles of the quantities $\phi_0$ and $\phi_i$, respectively:
\begin{equation}
\label{eq:phi_mu_roles}
dx^0 = \frac{1}{2}\left[\frac{dx_0^{(2)}}{\left(1 + \frac{\phi_0}{g_0}  \right)} + \frac{dx_0^{(1)}}{\left(1 + \frac{\phi_0}{g_0}  \right)}\right] - \frac{\phi_i}{g_0\left(1 + \frac{\phi_0}{g_0}  \right)} dx^i.
\end{equation}
Defining the scaled time intervals for light propagation in the two directions, $\tilde{dx_0}^{(A)}$, where $A=1,2$:
\begin{equation}
\label{scaled_times}
\tilde{dx_0}^{(A)} = \frac{dx_0^{(A)}}{\left(1 + \frac{\phi_0}{g_0}  \right)}, 
 \end{equation}
we can rewrite Equation (\ref{eq:phi_mu_roles}) as:
\begin{equation}
\label{eq:phi_mu_roles2}
dx^0 = \frac{1}{2}\left(\tilde{dx_0}^{(2)} + \tilde{dx_0}^{(1)}\right) - \frac{\phi_i}{g_0\left(1 + \frac{\phi_0}{g_0}  \right)} dx^i.
\end{equation}
Therefore, the quantity $\phi_0$ scales the coordinate time interval for light to travel in each direction between $A$ and $B$.  In other words, $\phi_0$ modifies the local speed of light $\tilde{c}$:
\begin{equation}
\label{local_speed_light}
\tilde{c} = c\left(1 + \frac{\phi_0}{g_0}  \right).
\end{equation}
On the other hand, the quantities $\phi_i$ (with the appropriate $\phi_0$-dependent scaling) shift the definition of simultaneity in an analogous manner as the $g_i$ and define (along with the $g_i$) the difference in propagation speeds for light travel in each direction between $A$ and $B$.  Loosely stated, $\phi_0$ and $g_0$ define the two-way speed of light and $\phi_i$ and $g_i$ define the one-way speed of light.  This analysis reveals the space-time interpretation of the $\phi_\mu$ field. 

Note that the field $\phi_\mu$ does not transform as a four-vector, but the field $\psi_\mu \equiv g_\mu + \phi_\mu$ transforms as a four-vector so that the linear term $d\sigma$ is an invariant:
\begin{equation}
\label{eq:linear_element}
d\sigma = \psi_\mu dx^{\mu} = \left( g_\mu + \phi_\mu  \right) dx^{\mu}.
\end{equation}
Also note that the integrability of $d\sigma$ depends on the vanishing of the field:
\begin{equation}
\label{eq:skew_symmetric_tensor}
f_{\mu\nu} \equiv \psi_{\mu,\nu} - \psi_{\nu,\mu},
\end{equation}
where a comma denotes partial differentiation.  We see at once both a similarity and a difference between Equation (\ref{eq:skew_symmetric_tensor}) and the standard electromagnetic field tensor.  The integrability of the linear quantity (\ref{eq:linear_element}) is defined by a skew-symmetric tensor (\ref{eq:skew_symmetric_tensor}) that resembles the electromagnetic field tensor; however, the four-vector $\psi_\mu$ is not independent of the gravitational potentials.  Rather, the fundamental variable that is independent of the gravitational potentials is $\phi_\mu$, which exhibits vector transformation properties only under the restricted transformation (\ref{eq:restricted_transformations}).

\subsection{Example: Transformation to a Constant-Velocity Frame}
It is constructive to pause before developing the theory further in order to consider a simple example of the formalism developed above.  Therefore, we examine the transformation from a one-dimensional stationary frame to a frame moving with a constant relative velocity $v$.  In the rest frame the line element is:
\begin{equation}
ds^2 = dt^2 - dx^2,
\end{equation}
and the metric quantities that define physical time are:
\begin{eqnarray}
g_0 &=& 1 \nonumber \\
g_1 &=& 0 \nonumber \\
\phi_0 &=& 0 \nonumber \\
\phi_1 &=& 0
\end{eqnarray}
so that $d\sigma = \psi_\mu dx^\mu = dt$.
Let us now consider the transformation to a frame moving with relative velocity $v$:
\begin{eqnarray}
\label{galilean_trans}
x &=& x^{\prime} - v t^{\prime} \nonumber \\
t &=& t^{\prime}.
\end{eqnarray}
The line-element now becomes:
\begin{equation}
ds^2 = \left(1-v^2\right)dt^{\prime 2} + 2vdx^{\prime}dt^{\prime} - dx^{\prime 2},
\end{equation}
and the metric quantities that define physical time in the primed frame are:
\begin{eqnarray}
g_0^{\prime} &=& \sqrt{1-v^2} \nonumber \\
g_1^{\prime} &=& \frac{v}{\sqrt{1-v^2}} \nonumber \\
\phi_0^{\prime} &=& 1-\sqrt{1-v^2} \nonumber \\
\phi_1^{\prime} &=& -\frac{v}{\sqrt{1-v^2}}
\end{eqnarray}
so that $d\sigma = \psi^{\prime}_\mu dx^{\prime \mu} = dt^{\prime}$.  Note that $g_0+\phi_0 = g_0^{\prime} + \phi_0^{\prime} = 1$ and $g_1 + \phi_1 = g_1^{\prime} + \phi_1^{\prime} =0$.  The quantities $\phi_0^{\prime}$ and $\phi_1^{\prime}$ emerge in the primed frame to preserve $d\sigma$ under the transformation (\ref{galilean_trans}). 

Let us now examine simultaneity in the primed frame.  According to conventional relativity simultaneity is defined by the condition $g^{\prime}_\mu dx^{\prime \mu} =0$ so that:
\begin{equation}
\label{standard_relativity_sim}
dt^{\prime} = -\frac{v}{1-v^2}dx^{\prime},
\end{equation}
which results from the fact that the speed of light according to $dt^{\prime}$ is different in each direction between $A$ and $B$:
\begin{equation}
\frac{dx^{\prime}}{dt^{\prime}} = v\pm c.
\end{equation}
In traditional relativity simultaneity is defined to compensate for the difference in the one-way speeds of light.  On the other hand, according to the formalism developed above simultaneity is defined by the relationship $g^{\prime}_{\mu} dx^{\prime \mu} = -\phi_\mu dx^{\prime \mu}$:
\begin{equation}
\label{new_relativity}
dt^{\prime} = 0.
\end{equation}
At first glance, it appears that the proposed theory conflicts with special relativity.  However, no contradiction exists because simultaneity in the new theory is defined with a local speed of light, $\tilde{c}$, different from $c$:
\begin{equation}
\label{new_light_speed}
\tilde{c} = \frac{c}{\sqrt{1-v^2}},
\end{equation}
that is the same for propagation in each direction between $A$ and $B$.  In traditional relativity, the difference in light speed in each direction is responsible for a nonzero temporal difference between the simultaneous events, namely $dt^{\prime} \neq 0$ (cf.\ Equation (\ref{standard_relativity_sim})).  In the reformulation of relativity proposed above, the quantities $\phi_i$ emerge so that the speed of light is the same in each direction between points $A$ and $B$.  This allows simultaneity to be defined by the vanishing of the coordinate time differential, $dt^{\prime}$, in the moving frame.

\section{\label{sec:field_equations}Field Equations}
The space-time interval of the proposed reformulation of general relativity is the same as that of the traditional theory (\ref{eq:metric1}) because the contribution of the new terms that results from the introduction of the field $\phi_\mu$ sums identically to zero in the space-time interval.  Therefore, the derivation of the new field equations will be similar to the derivation of the traditional gravitational field equations.  However, we will have to modify the derivation in order to satisfy the postulate of the preservation of the local properties of time.

The action integral of traditional general relativity is given by:
\begin{equation}
I = \int_D{\left( R-2\kappa L_F \right)\sqrt{-g}\,d^4x},
\end{equation}
where $R$ is the Ricci scalar curvature, $L_F$ describes all fields except the gravitational field, and $\kappa = 8\pi G$.  The field equations follow from the variational principle, $\delta I = 0$, where the $g_{\mu\nu}$ are varied independently subject only to the requirement that their variations $\delta g_{\mu\nu}$ as well as the variations of their first derivatives $\delta g_{\mu\nu,\lambda}$ vanish on the boundary of integration.  This yields the well-known Einstein field equations:
\begin{equation}
G_{\mu\nu} = \kappa T_{\mu\nu},
\end{equation}
where $G_{\mu\nu}$ is the divergenceless Einstein tensor and $T_{\mu\nu}=-\frac{2}{\sqrt{-g}}\frac{\delta \left(L_F\sqrt{-g}\right)}{\delta g^{\mu\nu}}$ is the energy-momentum tensor of all the other fields. 

As stated above, we must modify the derivation in order to preserve the local properties of time under the variations $g_{\mu\nu}$.  This can be accomplished with the method of Lagrange multipliers.  First, we note that the integrability of $d\sigma$ is determined by the antisymmetric tensor $f_{\mu\nu} \equiv \psi_{\mu,\nu} - \psi_{\nu,\mu}$.  Consequently, the quantity $f_{\mu\nu}f^{\mu\nu}$ is an invariant characterization of the local properties of time, which must be preserved under the variations of $g_{\mu\nu}$.  Therefore, we introduce the constraint:
\begin{equation}
\label{eq:constraint}
f_{\mu\nu}f^{\mu\nu} = f_0,
\end{equation}
into the action of the gravitational field:
\begin{equation}
I_G = \int_D{\left( R-2\kappa L_F \right)\sqrt{-g}\,d^4x} + \lambda_0\int_D{\left(f_{\mu\nu}f^{\mu\nu} - f_0\right)\sqrt{-g}\,d^4x},
\end{equation}
where $f_0$ is externally defined and $\lambda_0$ is a dimensionless Lagrange multiplier field.  We vary the action with respect to the variables $g_{\mu\nu}$ and $\phi_\mu$\footnote{Note that we can alternatively take $\psi_\mu$ as an independent quantity for the variation since the $g_{\mu\nu}$ are to be held fixed during this variation.}.  Variation of the action with respect to the Lagrange multiplier gives the constraint (\ref{eq:constraint}).  Variation of the action with respect to the fields $g_{\mu\nu}$ and $\phi_\mu$ gives:
\begin{eqnarray}
\label{new_field_equations}
G_{\mu\nu}+\Lambda g_{\mu\nu} + \kappa \Omega_{\mu\nu} &=& \kappa T_{\mu\nu} \nonumber \\
f^{\mu\nu}_{\;\; ;\nu} &=& j^\mu, 
\end{eqnarray}
where $j^{\mu}=\frac{2\kappa}{\lambda_0\sqrt{-g}}\frac{\delta \left(L_F\sqrt{-g}\right)}{\delta \phi_{\mu}}$, a semicolon denotes covariant derivative, $\Lambda \equiv \frac{\lambda_0 f_0}{2}$, and
\begin{equation}
\Omega_{\mu\nu} \equiv \frac{2\lambda_0}{\kappa} \left(T_{\mu\nu}^{(\psi)} + t_{\mu\nu} \right),
\end{equation}
with
\begin{equation}
\label{psi_tensor}
T_{\mu\nu}^{(\psi)} \equiv  \left(f_{\mu\alpha}f_{\nu}^{\;\alpha} - \frac{1}{4}f_{\alpha\beta}f^{\alpha\beta}g_{\mu\nu}  \right)
\end{equation}
and 
\begin{eqnarray}
t^{00} &\equiv& \left[g_{00}^{-1/2}f^{0\nu} - g_{0i}g_{00}^{-3/2}f^{i\nu} \right]_{,\nu} -\left[-\frac{1}{2}g_{00}^{-3/2}g_{00,\nu}f^{0\nu} + \frac{3}{2}g_{0i}g_{00}^{-5/2}g_{00,\nu}f^{i\nu} - g_{00}^{-3/2}g_{0i,\nu}f^{i\nu}\right]  \nonumber \\
&=& g^{-1/2}_{00}f^{0\nu}_{\;\;,\nu} - g_{0i}g^{-3/2}_{00}f^{i\nu}_{\;\;,\nu} \nonumber \\
t^{0i} &\equiv& \left[2g_{00}^{-1/2}f^{i\nu}    \right]_{,\nu} + g_{00}^{-3/2}g_{00,\nu}f^{i\nu} = 2g_{00}^{-1/2}f^{i\nu}_{\;\;,\nu}\nonumber \\
t^{ij} &\equiv& 0.
\end{eqnarray}  
Note that $j^{\mu}$ and $f_0$ cannot be prescribed arbitrarily.  The source term $j^{\mu}$ must be defined so that the solution of $f^{\mu\nu}_{\;\; ;\nu} = j^\mu$ satisfies the constraint $f_{\mu\nu}f^{\mu\nu} = f_0$.

We see that a `cosmological constant' term emerges naturally in this theory and is proportional to the quantity $f_0$ in the constraint.  By substituting the constraint (\ref{eq:constraint}) in (\ref{new_field_equations}), the `cosmological constant' term cancels out, producing the following field equations:
\begin{equation}
\label{new_field_equations2}
G_{\mu\nu}+ 2\lambda_0 \left(f_{\mu\alpha}f_{\nu}^{\;\alpha} + t_{\mu\nu} \right)= \kappa T_{\mu\nu};
\end{equation}
By taking the covariant divergence of the above equations we obtain an equation for the Lagrange multiplier field:
\begin{equation}
\left[2\lambda_0 \left(f^{\mu}_{\;\;\alpha}f^{\nu\alpha} + t^{\mu\nu}\right)\right]_{;\nu}=\kappa T^{\mu\nu}_{\;\; ;\nu}.
\end{equation} 
Also, we see that $f^{\mu\nu}$ satisfies Maxwell's equations and the tensor $T^{(\psi)}_{\mu\nu}$ in Equation (\ref{psi_tensor}) resembles the standard electromagnetic stress-energy tensor:
\begin{equation}
\label{em_tensor}
T_{\mu\nu}^{\textrm{(EM)}} \equiv  \frac{1}{4\pi}\left(F_{\mu\alpha}F_{\nu}^{\;\alpha} - \frac{1}{4}F_{\alpha\beta}F^{\alpha\beta}g_{\mu\nu}  \right),
\end{equation}
where $F_{\mu\nu} = A_{\mu,\nu}-A_{\nu,\mu}$ and $A_\mu$ are the electromagnetic potentials of the traditional Einstein-Maxwell theory.  Therefore, we identify the field $\psi_\mu$ with the classical electromagnetic field such that:  
\begin{equation}
\label{eq:EM_field_identification}
\psi_\mu = \alpha A_\mu,
\end{equation}
where $\alpha = \sqrt{\frac{\kappa}{8\pi\lambda_0}}$.  Note that $\psi_\mu$ (and hence $A_\mu$) is composed of both $g_{\mu}$ and $\phi_\mu$.  Therefore, the generally covariant electromagnetic field is composed of terms from both the standard gravitational tensor $g_{\mu\nu}$ as well as the new degrees of freedom $\phi_\mu$.  Note also that $\phi_\mu = \alpha A_\mu$ when $g_\mu = 0$.  

\section{\label{sec:equations_motion}Equations of Motion} 

In traditional general relativity, a point-like particle moving between two points $A$ and $B$ in Riemannian space traverses a geodesic.  The equations of motion follow from the variational principle:
\begin{equation}
\label{variational_principle}
\delta\int_A^B{m\,ds} = \delta\int_A^B{L_0\,d\tau}=0,
\end{equation}
where $m$ is the mass of the particle and $L_0=m\sqrt{g_{\mu\nu}\frac{dx^\mu}{d\tau}\frac{dx^\nu}{d\tau}}$ is the free-particle Lagrangian.  The variation is made arbitrarily, subject only to the constraint of fixed endpoints, producing:
\begin{equation}
\label{geodesicequations1}
\frac{d^2 x^\mu}{d\tau^2} + \Gamma^{\mu}_{\alpha\beta}\frac{dx^\alpha}{d\tau}\frac{dx^\beta}{d\tau}
=0.
\end{equation}

Since $ds$ is varied arbitrarily and is only subject to the endpoint constraints, the quantity $d\sigma = \psi_\mu dx^\mu$ is not preserved under the variations.  Consequently, the standard variational principle (\ref{variational_principle}) for geodesics does not satisfy the postulate of the preservation of the local properties of time.  Therefore, we must reformulate the variational principle in order to preserve physical time under the variations of the space-time path.  To this end, we introduce the local constraint along the parameterized space-time path:
\begin{equation}
\label{sigma_constraint}
d\sigma  = d\sigma_0,
\end{equation}
where $d\sigma_0$ defines physical time intervals at every point on the space-time path.  Note that $d\sigma$ need not be integrable.  Using the method of Lagrange multipliers, we write the new Lagrangian as:
\begin{equation}
\label{lagrangian2}
L=L_0 + \lambda_1 \left( \psi_\mu\frac{dx^\mu}{d\tau} - \frac{d\sigma_0}{d\tau}\right),
\end{equation}
where $\lambda_1$ is the Lagrange multiplier that also absorbs the constant of proportionality so that each term has the same units.  Note that $\lambda_1$ must be a constant so that the new action is invariant.  Variation of the Lagrange multiplier yields the constraint (\ref{sigma_constraint}), whereas the variations of the parameterized space-time path produce a new term in the equations of motion that acts as a Lorentz force:
\begin{equation}
\label{geodesicequations2}
\frac{d^2 x^\mu}{d\tau^2} + \Gamma^{\mu}_{\alpha\beta}\frac{dx^\alpha}{d\tau}\frac{dx^\beta}{d\tau} = \frac{\lambda_1}{m} f^{\mu}_{\,\,\nu}\frac{dx^\nu}{d\tau},
\end{equation}
if we identify $\lambda_1$ as being proportional to electric charge.  Therefore, the Lorentz force acquires a space-time interpretation: A charged particle deviates from geodesic motion in order to satisfy the local constraints imposed on physical time along the trajectory.  Note that the quantity $f_{\mu\nu}$ is responsible for the Lorentz force and therefore the emergence of the field $\phi_\mu$ alone is not sufficient to produce an electromagnetic deflection.  For example, consider the emergence of $\phi_\mu$ as a result of a coordinate transformation from a frame in which $f_{\mu\nu} = 0$.  An example of this type of field is the $\phi_\mu$ field that emerges as a result of a coordinate transformation from the laboratory frame to the rotating frame.  Since the condition $f_{\mu\nu} = 0$ is preserved under an arbitrary coordinate transformation, then according to Equation (\ref{geodesicequations2}), this field will have no effect on the motion of charged particles.  On the other hand, when $f_{\mu\nu} \neq 0$ (non-integrable $d\sigma$) charged particles will deflect according to the Lorentz force term in Equation (\ref{geodesicequations2}). 

\section{Weak Field Limit}
In this section we consider the weak-field approximation \cite{misnerthornewheeler} of the new set of field equations (\ref{new_field_equations}).  In this approximation the metric may be written as a linear perturbation from the flat space-time metric, $\eta_{\mu\nu}$:
\begin{equation}
g_{\mu\nu} = \eta_{\mu\nu} + h_{\mu\nu},
\end{equation} 
where $\left|h_{\mu\nu}\right|\ll 1$ is the small perturbation.  The field equations for the $f^{\mu\nu}$ become:  
\begin{equation}
\label{linearized_maxwell}
f^{\mu\nu}_{\;\; ,\nu}=j^\mu.
\end{equation}
We solve for the metric terms in a region outside the source $j^{\mu}$ so that in the region of interest:
\begin{equation}
\label{linearized_maxwell_no_source}
f^{\mu\nu}_{\;\; ,\nu}=0.
\end{equation} 
and therefore the terms $t^{\mu\nu}$ vanish.  As a result, the remaining field equations in the `Lorentz gauge' ($\bar{h}^{\mu\nu}_{\;\;,\nu}=0$) are:
\begin{equation}
\label{linearized_einstein}
\Box \bar{h}_{\mu\nu} = -2\kappa \left( T_{\mu\nu} - \frac{2\lambda_0}{\kappa}f_{\mu\alpha}f_{\nu}^{\;\alpha}  \right),
\end{equation}
where $\Box = \partial_\mu \partial^\mu$ and $\bar{h}_{\mu\nu}\equiv h_{\mu\nu} - \frac{1}{2}\eta_{\mu\nu}h$, with $h=\eta^{\mu\nu}h_{\mu\nu}$.  Therefore, in the weak-field limit the equations of the proposed reformulation of general relativity are very similar to the equations of general relativity in the presence of an electromagnetic field; this is a result of the vanishing of the $t^{\mu\nu}$ in regions for which $j^{\mu}=0$.  The difference in the two theories, however, should lead to measurably different predictions.

We follow the usual treatment and assume that the perturbation $h_{\mu\nu}$ is expressed in isotropic space coordinates so that $h_s=h_{11}=h_{22}=h_{33} $ and that the matter is slowly moving with low density and negligible pressure:
\begin{equation}
T^{\mu\nu} = \rho u^\mu u^\nu,
\end{equation}
where $u^\mu$ is the 4-velocity and $\rho$ is the matter density.  Therefore, the linearized field equations (\ref{linearized_einstein}) become:
\begin{eqnarray}
\label{linearized_einstein2}
\Box \Phi &=& -\frac{\kappa}{2}\left(\rho - \frac{2\lambda_0}{\kappa}f_{0\alpha}f_{0}^{\;\alpha} \right)\nonumber \\
\Box \vec{h} &=& 2\kappa \rho\left(\vec{v} + \frac{2\lambda_0}{\kappa \rho}\vec{f}^{(\psi)}   \right),
\end{eqnarray} 
where $h_{00} = h_s = 2\Phi$, $\vec{h}=\left\{h_{01},h_{02},h_{03}\right\}$, $\vec{v}$ is the velocity of the source, and $\vec{f}^{(\psi)}= \left\{f_{0\alpha}f_{1}^{\;\alpha},f_{0\alpha}f_{2}^{\;\alpha},f_{0\alpha}f_{3}^{\;\alpha}\right\}$.  When $f_{0\alpha}f_{0}^{\;\alpha}=0$ and $\vec{f}^{(\psi)}=0$  these equations reduce to the standard linearized gravitational equations.  In this case the equations may be solved to produce (neglecting retardation effects):
\begin{eqnarray}
\label{standard_solutions}
\Phi\left(\vec{r}\right) &=& - G\int\frac{\rho\left(\vec{r}^{\prime}\right)\,d^3\vec{r}^{\prime}}{\left|\vec{r} -  \vec{r}^{\prime} \right|}         \nonumber \\
 \vec{h}\left(\vec{r}\right) &=&  4G\int\frac{\rho\left(\vec{r}^{\prime}\right)\vec{v}\left(\vec{r}^{\prime}\right)\,d^3\vec{r}^{\prime}}{\left|\vec{r} -  \vec{r}^{\prime} \right|}. 
\end{eqnarray}
This results in the well-known Lense-Thirring line element:
\begin{equation}
\label{lense_thirring}
ds^2 = \left( 1 + 2\Phi \right) dt^2 - \left(1 - 2\Phi  \right)\vec{dr^2} + 2\vec{h}\cdot \vec{dr}dt.
\end{equation}
However, when $f_{0\alpha}f_{0}^{\;\alpha}\neq 0$ or $\vec{f}^{(\psi)}\neq0$ one must recognize the additional dependence on the metric that emerges in Equation (\ref{linearized_einstein2}).  In this case the solutions of Equations (\ref{linearized_einstein2}) are not given by Equations (\ref{standard_solutions}).  We immediately see that the above reformulation of general relativity suggests that the Lense-Thirring effect in the presence of an electromagnetic field will differ slightly from that predicted by standard general relativity.
  
\section{Implications for Outstanding Problems}
In this section we discuss the implications of our reformulation of general relativity on some of the aforementioned outstanding problems.  

\subsection{The Pioneer Anomaly}

As is well known, analyses of radio Doppler and ranging data from the Pioneer missions indicate that there is an anomalous blueshift in the detected microwave signal, which can either be attributed to an anomalous acceleration of the spacecraft $a_P\sim 8.7\times 10^{-8} \mathrm{cm}\,\mathrm{s}^{-2}$ directed towards the sun or to an acceleration of clocks $a_t = a_P/c \sim 2.9 \times 10^{-18} \textrm{s}^{-1}$ \cite{andersonetal1988}.  While Page et al.\ \cite{pageetal2009} have recently argued that the current ephemeris of Pluto does not preclude the existence of the Pioneer effect, it is unlikely that such an acceleration, which is four orders of magnitude greater than the largest relativistic corrections to Newtonian gravity, has a gravitational origin \cite{iorio2007}.  Such an acceleration would require a violation of the weak equivalence principle at the outer radii of the Solar System.  Therefore, if the anomalous blueshift is a result of unexplained physics and not a systematic error, then it is most likely due to a clock acceleration that does not manifest itself in a physical acceleration of this magnitude in our solar system.  In this section, we show how the proposed reformulation of gravity naturally accounts for this anomaly.

We start by writing the Schwarzschild line-element, which is also a solution of the free-field field equations in the proposed reformulation of general relativity:
\begin{equation}
\label{schwarzschild}
ds^2 = \left(1-\frac{2MG}{r}\right)dt^2 - \left(1-\frac{2MG}{r}\right)^{-1}dr^2-r^2d\Omega^2,
\end{equation}
where $M=2\times 10^{33}\,\textrm{gm}$ is the mass of the sun.  We assume all quantities are independent of time and are spherically symmetric.  According to Equation (\ref{eq:EM_field_identification}) there will be an electric field present due to the radial dependence of the quantity $\psi_0$:  
\begin{equation}
\psi_0 = g_0 + \phi_0 \approx 1 - \frac{MG}{r} + \phi_0.
\end{equation}
We first consider the case where the electric field at the spacecraft vanishes.  In this case, $\psi_0$ must be constant; therefore we conclude $\phi_0 = \frac{MG}{r} + a_0$, where $a_0$ is an arbitrary constant, which can be set to zero.

According to Equation (\ref{local_speed_light}) the local speed of light at the location of the spacecraft is: 
\begin{equation}
\label{local_speed_light_pioneer}
\tilde{c} = c\left(1 + \frac{\phi_0}{g_0}  \right) \simeq c\left( 1 + \frac{MG}{c^2r} \right),
\end{equation}
where we have now introduced $c$ explicitly back into all the terms.  Therefore, according to the definition of simultaneity in general relativity, the following distance for light travel is neglected: 
\begin{equation}
dL = \frac{MG}{cr}dt,
\end{equation}
which will result in attributing an apparent acceleration to the spacecraft:
\begin{equation}
\label{anomalous_acceleration}
\frac{d^2L}{dt^2} = -\frac{MGv}{cr^2},
\end{equation}
where $v$ is the velocity of the spacecraft.  Using $v=12\, \textrm{km}\,\textrm{s}^{-1}$ and $r=20\, \textrm{AU}$, we obtain for the `anomalous acceleration':
\begin{equation}
\frac{d^2L}{dt^2} = - 5.9\times 10^{-8} \,\textrm{cm}\,\textrm{s}^{-2},
\end{equation}
which is to be compared to the anomalous Pioneer acceleration $a_P\sim -8.7\times 10^{-8} \mathrm{cm}\,\mathrm{s}^{-2}$.  Note that the acceleration derived above is not a `real' acceleration of the spacecraft, but is only apparent due to the incorrect speed of light used in the equations of traditional general relativity.  We can derive the same result by considering the acceleration of the clocks at the location of the spacecraft (cf.\ Equation (\ref{scaled_times})):
\begin{equation}
\tilde{dt}^{(\textrm{Pioneer})} = \frac{dt}{\left(1 + \frac{\phi_0}{g_0}  \right)} \simeq dt\left(1-\phi_0\right), 
\end{equation}
so that the clocks on the Pioneer spacecraft are accelerated relative to the clocks of traditional relativity according to:
\begin{equation}
\frac{d^2\tilde{t}^{(\textrm{Pioneer})}}{dt^2} = \frac{MGv}{c^2r^2} = \frac{1}{c}\left|\frac{d^2L}{dt^2}\right|.
\end{equation}

While the simple model presented above provides encouraging results at $20\,\textrm{AU}$, it predicts a $r^{-2}$ dependence of the anomalous acceleration that is not observed in the data.  For example, Equation (\ref{anomalous_acceleration}) predicts an anomalous acceleration of $- 1.48\times 10^{-8} \,\textrm{cm}\,\textrm{s}^{-2}$ at $40 \,\text{AU}$.  On the other hand, the Pioneer's anomalous acceleration has been observed to be approximately constant for radii $20 - 70 \,\textrm{AU}$.  This radial dependence can be traced back to the assumption that the electromagnetic field vanishes at the radii of interest.  A more complete analysis requires a model for the electromagnetic field at radii $20 - 70 \,\textrm{AU}$.

\subsection{The Dark Matter Problem}
First discovered by Zwicky in the 1930's \cite{zwicky1933,zwicky1937}, velocities on the galactic scale are much larger than those predicted by general relativity when the source of the gravitational field is taken to be the observed visible matter (see also, e.g., \cite{einastoetal} and \cite{rubinetal}).  Zwicky postulated, and it is now generally accepted, that a considerable amount of non-visible matter must be present in the extragalactic regime in order to provide the additional acceleration required to maintain these excessive velocities.  This non-visible matter is commonly called dark matter and is believed to resolve acceleration discrepancies observed in systems ranging from dwarf spheroidal galaxies with visible masses $\sim 10^7 M_{\odot}$ to clusters of galaxies with observed masses $\sim 10^{14} M_{\odot}$.  Furthermore, dark matter is believed to play a key role in structure formation of the universe and primordial nucleosynthesis, and is believed to significantly affect the anisotropy of the cosmic microwave background.  Excellent reviews of the dark matter problem are given in Refs. \cite{turner,silk}.

Despite thirty years of laboratory experiments and astronomical observation, dark matter has never been observed directly \cite{Bertoneetal}; its existence is only inferred indirectly due to its purported gravitational effects on visible matter.  Modifications of gravitational theory have been proposed \cite{finzi,tohline,sanders1984,sanders1986,goldman,kuhnandkruglyak} that may eliminate the need for dark matter, and perhaps the most well-known is the modified Newtonian dynamics (MOND) theory \cite{milgrom1983a,milgrom1983b,milgrom1983c}.  MOND is characterized by an acceleration scale and predicts departures from a Newtonian force law in the extragalactic regime where dynamical accelerations are small.  Recently, a relativistic generalization of MOND, Tensor-Vector-Scalar (TeVeS), was proposed  \cite{bekenstein2004} that resolves some of the earlier problems of the MOND theory.  However, TeVeS has not been experimentally confirmed.  In this section we show how the proposed reformulation of gravity naturally accounts for the `anomalous' acceleration observed in rotating spiral galaxies, without the need to postulate dark matter.  

We model a galaxy rotating in the azimuth ($\hat{\phi}$) direction with the Schwarzschild line element (\ref{schwarzschild}) at radius $r \gg r_0$, where $r_0$ is the effective radius of the visible matter distribution; the quantity $M$ now refers to the total visible mass of the galaxy contained within $r<r_0$.  Note that we are assuming the weak-field frame-dragging term, $g_3 \simeq 2GJ/r^2$, where $J$ is the angular momentum of the galaxy, is negligible at $r \gg r_0$.  According to the proposed reformulation of general relativity, the electromagnetic field is a result of the coordinate-dependence of $\psi_\mu = g_\mu + \phi_\mu$.  Since we are assuming the $g_i$ are negligible ($r\gg r_0$) we can write $\psi_i \simeq \phi_i$.  

We are particularly interested in the quantity $\phi_3$, since according to Equation (\ref{eq:phi_mu_roles2}) this will modify the one-way speed of light in the azimuth ($\hat{\phi}$) direction - in exactly the same way the quantities $g_i$ are responsible for a difference in the one-way and two-way speeds of light in the conventional theory.  In other words, traditional general relativity will incorrectly attribute an additional velocity $\phi_3$ to the rotational motion of the galaxy when inferring velocity from Doppler shifts of electromagnetic signals.  At the tail-end of the galactic rotation curve ($r\gg r_0$) this additional velocity is approximately constant.  Therefore, at $r\gg r_0$ we assume $\phi_3 = A_0$, where $A_0$ is a constant.  As a result, $\psi_3 = g_3 + \phi_3 \simeq \phi_3 = A_0$.  In the plane of the galaxy, this corresponds to a perpendicular magnetic field with a $r^{-1}$ dependence:
\begin{equation}
B_\theta = -\frac{cA_0}{\alpha r}.
\end{equation}
As $r\rightarrow r_0$, one expects $\phi_3 \neq \textrm{constant}$ because the frame-dragging term, $g_3$, cannot be considered negligible at these radii.  If we assume the magnetic field retains its same functional form as $r\rightarrow r_0$, then $\phi_3$ would need to be:
\begin{equation}
\phi_3 = A_0 - 2GJ/r^2,
\end{equation}
in order to compensate for the $g_3$ term.  A more complete analysis will require a detailed model of the galactic magnetic field as well as a better model of the galactic gravitational field \cite{neugebauermeinel1,neugebauermeinel2,neugebauermeinel3}.  Nevertheless, we observe how the presence of a galactic electromagnetic field can modify the local one-way speed of light and lead to seemingly anomalous measurements of the galactic rotational velocities.  It should be emphasized that this difference in velocity is not due to a real acceleration of the rotating matter, but is due to an incorrect quantification of the azimuthal one-way speed of light in traditional general relativity.  Hence, there is no need to postulate dark matter to account for the anomalous acceleration observed in rotating spiral galaxies.  

\subsection{The Dark Energy Problem}
The Friedmann-Lema\^itre $\Lambda\textrm{CDM}$ model of cosmology has been accepted by the scientific community as the new Standard Model of Cosmology \cite{steinhardtostriker,bahcalletal}.  It supercedes the previous Standard Model of Cosmology, embracing all of its accomplishments, and claims additional success.  This model agrees closely with a wide range of observations, including measurements of the abundance of primordial elements, CMB anisotropies, the age of the universe, the luminosity of supernovae, and the large scale structure of the universe.  According to the $\Lambda\textrm{CDM}$ model, the universe is spatially flat and was initiated with the Big Bang, a state of infinite density and temperature, approximately $15 \times 10^9$ years ago.  This was followed by a potential, or vacuum, energy-dominated (inflation) phase, a radiation-dominated phase, and a matter-dominated phase.  It is believed that the universe is presently transitioning from the matter-dominated phase to a cosmological constant-dominated phase.

First reported in Refs. \cite{perlmutteretal, riess1, riess2}, observations of Type Ia Supernova (SN Ia) indicate that the universe is accelerating.  The $\Lambda\textrm{CDM}$ model attributes this acceleration to the cosmological constant, $\Lambda$, which was originally introduced into general relativity by Einstein \cite{einstein1917} in order to permit homogeneous, static solutions of the field equations.  However, the introduction of the cosmological constant brings a number of problems in its wake, including the well known cosmological constant, or fine-tuning, problem.  This results from the observation that the contribution to the vacuum energy density from quantum fields behaves like a cosmological constant, and is according to modern particle theories orders of magnitude larger than the measured cosmological constant, which is crudely approximated by $\Lambda \approx H_0^2$, where $H_0$ is the present value of the Hubble parameter.  Consequently, considerable effort is being exerted to replace $\Lambda$ in the $\Lambda\textrm{CDM}$ model with more general forms of dark energy that are typically described by scalar fields such as quintessence, K-essence, tachyon, phantom and dilatonic models (see \cite{copelandetal} for an excellent review). 

We have seen above that the cosmological constant emerges in the proposed reformulation of general relativity and because of the constraint (\ref{eq:constraint}) it cancels out of the field equations.  Therefore, a deeper understanding of the constraint and its relationship with quantum fields can shed light on the remarkable vanishing of the various contributions to the vacuum energy.  In other words, the vanishing of the various contributions to the vacuum energy may be attributed to a symmetry principle:  all contributions to the energy density are subject to the requirement that they preserve the local properties of time in the universe.  Thus, the $\psi_\mu$ field, along with the other quantum fields,
are subject to the constraint, $f_0 = f_{\mu\nu}f^{\mu\nu}$, which guarantees the vanishing of the cosmological constant.

At first glance, a theory that predicts a vanishing cosmological constant may seem to contradict the supernova observations.  However, this is not the case because the proposed theory (even with a vanishing cosmological constant) can account for the apparent acceleration of the universe.  Let us consider the general solution to the field equations (\ref{new_field_equations}) for a homogeneous and isotropic cosmological model with vanishing electromagnetic field ($f_{\mu\nu}=0)$:
\begin{eqnarray}
ds^2 &=& dt^2 - a(t)^2 d\Sigma^2 \nonumber \\
\phi_0 &=& \phi_0(t),
\end{eqnarray}
where $ds^2$ is the Robertson-Walker line element, $d\Sigma^2$ represents the three-dimensional line element of uniform curvature, $a(t)$ is the scale factor, and $\phi_0(t)$ is an arbitrary function of time.  Note that a vanishing electromagnetic field does not force the $\phi_\mu$ field to vanish entirely; rather, a general homogeneous and isotropic solution with vanishing electromagnetic field admits $\phi_0=\phi_0(t)$.  Therefore, the general cosmological model predicted by the proposed reformulation of general relativity naturally includes a speed of light that varies with time:
\begin{equation}
\label{cosmological_speed_light}
c(t) = c\left(1 + \frac{\phi_0(t)}{g_0}  \right) = c\left(1 + \phi_0(t)  \right).
\end{equation}
Such a variation of the speed of light can account for the apparent acceleration of the expansion of the universe \cite{sanejouand2009}, without requiring the introduction of a cosmological constant.  Note that the proposed theory shares similarities with Variable Speed of Light (VSL) theories \cite{magueijo2003}, and therefore, it can also provide insight into other cosmological problems, such as the horizon, flatness, homogeneity and isotropy problems \cite{albrechtMagueijo1999}.

\subsection{Quantum Gravity}
The proposed reformulation of general relativity provides an invariant definition of time and consequently may shed light on the `problem of time' in quantum gravity \cite{isham1993}.  Therefore, it is constructive to consider a quantum theory of gravity based on the theory presented above.  While it is out of the scope of the present paper to develop a complete theory of quantum gravity we discuss some of its salient features in this section.

Our starting point is the ADM formulation of general relativity \cite{ADM1962}, which is based on the following decomposition of the metric tensor:
\begin{equation}\left(g_{\mu\nu} \right) = \left( \begin{array}{cc}
-N^2 +N_iN^i & N_j  \\
N_i & h_{ij} \end{array} \right)
\end{equation}
and
\begin{equation}
\left(g^{\mu\nu} \right) = \left( \begin{array}{cc}
-N^{-2}  & N^{-2}N^j  \\
N^{-2}N^i & h^{ij}-N^{-2}N^iN^j \end{array} \right),
\end{equation}
where $N$ and $N_i$ are commonly known as the lapse and shift vector, respectively, $h_{ik}$ is the induced 3-metric on a hypersurface $\Sigma(t)$ at constant $t$, and the following relations hold: $h_{ik}h^{kj}=\delta_i^{\;j}$, $N^i = h^{ij}N_j$, $-g = N^2 h$, and $h=\textrm{det}\left(h_{ij}\right)$.  The traditional gravitational Lagrangian, $R\sqrt{-g}$, can be written in terms of the canonical variable set $\left\{N, N_i, h_{ij}\right\}$:
\begin{equation}
\label{ADM_lagrangian}
\mathcal{L}_\textrm{ADM} = Nh^{1/2}\left({}^{(3)}R+K_{ij}K^{ij} - K^2 \right),
\end{equation}
where $K_{ij}=\frac{1}{2}N^{-1}\left(N_{i.j}+N_{j.i}-h_{ij,0}  \right)$ is the second fundamental form, $K^{ij}=h^{ik}h^{jl}K_{kl}$, $K=h^{ij}K_{ij}$ and dots denote covariant differentiation based on the 3-metric $h_{ij}$.  
The canonical momenta corresponding to the variables $\left\{N, N_i, h_{ij}\right\}$ are:
\begin{eqnarray}
\label{canonical_momenta}
\pi &=& \frac{\partial \mathcal{L}_\textrm{ADM}}{\partial N_{,0}} = 0 \nonumber \\
\pi^{i} &=& \frac{\partial \mathcal{L}_\textrm{ADM}}{\partial N_{i,0}} = 0 \nonumber \\
\pi^{ij} &=& \frac{\partial \mathcal{L}_\textrm{ADM}}{\partial h_{ij,0}} = -\sqrt{h}\left(K^{ij} - h^{ij}K\right).
\end{eqnarray}
The first two equations are known as primary constraints; the momenta $\pi$ and $\pi^i$ vanish because the Lagrangian is independent of the velocities $N_{,0}$ and $N_{i,0}$.  The Hamiltonian is calculated in the usual way:
\begin{eqnarray}
\label{ADM_hamiltonian}
\mathcal{H}_\textrm{ADM} &=& \pi N_{,0} + \pi^iN_{i,0} + \pi^{ij}h_{ij,0} - \mathcal{L}_{ADM}  \nonumber \\
\mathcal{H}_\textrm{ADM} &=& \pi N_{,0} + \pi^iN_{i,0} + N\mathcal{H} + N_i\chi^i 
\end{eqnarray}
where $\mathcal{H} = h^{1/2}\left(K_{ij}K^{ij} - K^2-  {}^{(3)}R\right)$ and $\chi^i = -2\pi^{ij}_{\;\;,j} - h^{il}\left(2h_{jl,k}-h_{jk,l} \right) \pi^{jk}$.  Since $\pi$ and $\pi^i$ vanish, the ADM Hamiltonian can be written:
\begin{equation}
\label{ADM_hamiltonian2}
\mathcal{H}_\textrm{ADM} =  N\mathcal{H} + N_i\chi^i. 
\end{equation}
It is straightforward to derive Einstein's free-field equations by taking the Poisson bracket of the dynamical variables $\left\{N, N_i, h_{ij}\right\}$ with the Hamiltonian (\ref{ADM_hamiltonian2}) and enforcing the primary constraints.
 
By taking the Poisson bracket of the primary constraints with the Hamiltonian, one obtains the secondary constraints:
\begin{eqnarray}
\label{secondary_constraints_traditional}
\mathcal{H} &=& 0 \nonumber \\
\chi^i &=& 0.
\end{eqnarray}
These secondary constraints restrict the dynamics in order to preserve the primary constraint equations for all time.  These equations are often called the Hamiltonian constraint and the momentum constraint, respectively.  We see that the ADM Hamiltonian is the sum of the secondary constraints, with arbitrary Lagrange multipliers $N$ and $N_i$, respectively.  

The Wheeler-Dewitt theory of quantum gravity follows by elevating the Poisson brackets to commutators and turning the constraints into conditions on the state vector $\Psi$ \cite{dewitt1967}.  This leads to the following commutation relations: 
\begin{eqnarray}
\label{commutation_relations}
\left[N,\pi^{\prime}\right] &=& i\hbar\delta\left(\vec{x},\vec{x^{\prime}}\right) \nonumber \\
\left[N_i,\pi^{j^{\prime}}\right] &=& i\hbar\delta_i^{\;j^{\prime}} \nonumber \\
\left[h_{ij},\pi^{k^{\prime}l^{\prime}}\right] &=& i\hbar \delta_{ij}^{\;\;k^{\prime}l^{\prime}},
\end{eqnarray}
and the well-known Wheeler-DeWitt equations:
\begin{eqnarray}
\label{wheeler_dewitt}
\pi\Psi &=& 0 \nonumber \\
\pi^i\Psi &=& 0 \nonumber \\
\mathcal{H}\Psi &=& 0 \nonumber \\
\chi^i\Psi &=& 0,
\end{eqnarray}
where
\begin{eqnarray}
\delta_i^{\;j^{\prime}} &\equiv& \delta_i^{\;j} \delta\left(\vec{x},\vec{x}^{\prime} \right) \nonumber \\
\delta_{ij}^{\;\;k^{\prime}l^{\prime}} &\equiv&  \delta_{ij}^{\;\;kl} \delta\left(\vec{x},\vec{x}^{\prime} \right) \nonumber \\
\delta_{ij}^{\;\;kl} &\equiv& \frac{1}{2}\left(\delta_{i}^{\;k} \delta_{j}^{\;l} +\delta_{i}^{\;l} + \delta_{j}^{\;k} \right).
\end{eqnarray} 
As is well known, these equations are ill-defined.

We now turn to the equations of quantum gravity in the proposed reformulation of general relativity.  Because the dynamics are now constrained by the additional constraint (\ref{eq:constraint}), the Lagrangian of the new theory is:
\begin{equation}
\mathcal{L}^{\star} = \mathcal{L}_\textrm{ADM} + Nh^{1/2}\lambda_0\left(f_{\alpha\beta}f^{\alpha\beta}-f_0  \right).
\end{equation}
Note that the Lagrangian is no longer independent of the velocities $N_{,0}$ and $N_{i,0}$, and therefore, the canonical momenta (\ref{canonical_momenta}) now become:
\begin{eqnarray}
\label{canonical_momenta_new}
\pi &=& \frac{\partial \mathcal{L}}{\partial N_{,0}} = Nh^{1/2}\lambda_0\frac{\partial \left(f_{\mu\nu}f^{\mu\nu}\right)}{\partial N_{,0}} \nonumber \\
\pi^{i} &=& \frac{\partial \mathcal{L}}{\partial N_{i,0}} = Nh^{1/2}\lambda_0\frac{\partial \left(f_{\mu\nu}f^{\mu\nu}\right)}{\partial N_{i,0}} \nonumber \\
\pi^{ij} &=& \frac{\partial \mathcal{L}}{\partial h_{ij,0}} = -\sqrt{h}\left(K^{ij} - h^{ij}K\right) +Nh^{1/2}\lambda_0\frac{\partial \left(f_{\mu\nu}f^{\mu\nu}\right)}{\partial h_{ij,0}}
\end{eqnarray}
where the comma denotes a Lie-derivative in the direction of the time vector.  We treat $\lambda_0$ as a new canonical coordinate, which introduces the primary constraint:
\begin{equation}
\label{new_primary_constraint}
\pi_{\lambda_0} = \frac{\partial \mathcal{L}}{\partial \lambda_{0,0}} = 0.
\end{equation}
The new Hamiltonian is:
\begin{equation}
\mathcal{H}^{\star} = \pi N_{,0} + \pi^i N_{i,0} + \pi^{ij}h_{ij,0} + \pi_{\lambda_0}\lambda_{0,0}  - \mathcal{L}^{\star} .
\end{equation}
Note that the canonical momenta $\pi$ and $\pi^i$ no longer vanish and do not act as primary constraints; however, $\pi_{\lambda_0}$, the momentum conjugate to $\lambda_0$, does vanish.  By requiring that the Poisson bracket of the new primary constraint with the Hamiltonian $\mathcal{H}^{\star}$ vanishes, we obtain the secondary constraint:
\begin{equation}
f \equiv f_{\alpha\beta}f^{\alpha\beta}-f_0 =0.
\end{equation}
By taking the Poisson bracket of the secondary constraint $f$ with the Hamiltonian $\mathcal{H}^{\star}$, and iterating this process, one obtains a set of secondary constraints $\Phi_k$.
The Wheeler-Dewitt equations now become:
\begin{eqnarray}
\label{wheeler_dewitt_add}
\pi_{\lambda_0}\Psi &=& 0 \nonumber \\
\Phi_k\Psi &=& 0.
\end{eqnarray}
The commutation relations must now be formulated in terms of Dirac brackets instead of Poisson brackets \cite{dirac1964}:
\begin{equation}
\label{dirac_bracket_definition}
\left\{ f,g \right\}_{D} = \left\{ f,g \right\} - \left\{f,\Phi_k\right\}M^{-1}_{kl}\left\{\Phi_l,g\right\},
\end{equation}
where $\Phi_k$ are the second-class constraints and $M_{ij}=\left\{\Phi_i,\Phi_j\right\}$.  Therefore, the commutation relations are:
\begin{eqnarray}
\label{new_commutation_relations}
\left[\lambda_0,\pi_{\lambda_0}^{\prime}\right] &=& i\hbar\left\{\lambda_0,\pi_{\lambda_0}^{\prime}\right\}_{D} \nonumber \\
\left[N,\pi^{\prime}\right] &=& i\hbar\left\{N,\pi^{\prime}\right\}_{D} \nonumber \\
\left[N_i,\pi^{j^{\prime}}\right] &=& i\hbar\left\{N_i,\pi^{j^{\prime}}\right\}_{D} \nonumber \\
\left[h_{ij},\pi^{k^{\prime}l^{\prime}}\right] &=& i\hbar \left\{h_{ij},\pi^{k^{\prime}l^{\prime}}\right\}_{D}.
\end{eqnarray}

\section{\label{sec:discussion}Discussion}
We have argued that the time discontinuity paradox and the existence of CTCs in certain exact solutions of general relativity are manifestations of a fundamental a crisis in the foundations of Einstein's general relativity.  Therefore, we were forced to reformulate general relativity in a manner that is consistent with our physical experience of time.  Note that this conclusion is in sharp contradistinction to the widespread claim that the existence of CTCs in solutions to Einstein's equations forces one to accept the possibility of time travel.  We are arguing the converse, namely, that time travel has not been experimentally observed, and therefore CTCs (along with the time discontinuity) must be expunged from general relativity.  Additional reasons for revisiting these paradoxes include: `the problem of time' in quantum gravity and the observation that rotation is a general property of astrophysical systems exhibiting the dark matter problem.

We showed that we can resolve these paradoxes with a single postulate, namely, the preservation of the local properties of time.  As a result, the electromagnetic field emerges a `gauge' field that preserves the local properties of time in gravitational fields.  In this work, we introduced this postulate into general relativity by demanding $d\sigma$ remain a scalar invariant under arbitrary four-dimensional coordinate transformations.  This resulted in replacing the non-covariant definition of simultaneity in traditional general relativity with a covariant definition.  On the other hand, the postulate to preserve the local properties of time was introduced into the field equations by adding the constraint $f_0 = f_{\alpha\beta}f^{\alpha\beta}$ to the action.  While this constraint was introduced in order to prevent CTCs from emerging, we have not proven that this will be the case in general.  More work is needed to understand the role of CTCs in solutions of the new field equations.  If CTCs do arise in the proposed reformulation of general relativity, then one may need to introduce a stronger constraint into the variational principle.  For example, another invariant exists, $\star f_{\mu\nu}f^{\mu\nu}$, where $\star$ denotes dual, that can be preserved in the action principle.  Therefore, more work may be needed to refine the postulate of the preservation of the local properties of time in a way that forces a unique set of field equations consistent with our physical experience of time.

The proposed theory does not allow a clear separation of the gravitational and electromagnetic fields; the electromagnetic field $\psi_\mu$ (and hence $A_\mu$) is composed of both $g_{\mu}$ and $\phi_\mu$.  Therefore, the generally covariant electromagnetic field is composed of terms from the standard gravitational tensor, as well as the new degrees of freedom $\phi_\mu$.  Note that a general coordinate transformation can introduce the field $\phi_\mu$ into space-time, however, it cannot introduce non-integrability of $d\sigma$ (it cannot make $f_{\mu\nu}=0 \rightarrow f_{\mu\nu}\ne 0$).  For example, the $\phi_\mu$ field can emerge as a result of a coordinate transformation from the laboratory from to the rotating frame.  Since the condition $f_{\mu\nu} = 0$ is preserved under an arbitrary coordinate transformation, then according to Equation (\ref{geodesicequations2}), this field will have no effect on the motion of charged particles.  On the other hand, the usual electromagnetic source term $j_\mu\psi^\mu$, which must be consistent with the constraint $f_0 = f_{\alpha\beta}f^{\alpha\beta}$, can produce electromagnetic fields that deflect charged particles. 

A close examination of simultaneity revealed the space-time interpretation of the $\phi_\mu$ field.  The quantity $\phi_0$ modifies the local speed of light, whereas the quantities $\phi_i$ (with the appropriate $\phi_0$-dependent scaling) shift the definition of simultaneity and define (along with the $g_i$) the one-way speed of light.  Therefore, the proposed theory predicts variations of the speed of light in the presence of non-zero $\phi_\mu$, for both $f_{\mu\nu}=0$ and $f_{\mu\nu}\neq 0$.  In particular, the proposed theory predicts variations of the one-way and two-way speeds of light in the presence of a classical electromagnetic field.  Unfortunately, we cannot quantify the magnitude of such effects because the proportionality constant between the $\phi_\mu$ field and the classical electromagnetic field, $\alpha$, is proportional to the Lagrange multiplier $\lambda_0$.  Nevertheless, experimental verification of the proposed theory can be sought in measurements of the variation of the speed of light due to $\phi_0$ and Sagnac-like effects due to the quantities $\phi_i$, which can in turn provide an estimate of the Lagrange multiplier $\lambda_0$\footnote{Similarly, astrophysical measurements of the above predicted anomalous accelerations along with precise knowledge of the associated electromagnetic fields can provide an estimate of the Lagrange multiplier $\lambda_0$.}.  

There are numerous examples of astrophysical electromagnetic fields with unknown origin, ranging from the Earth's magnetic field to magnetic fields on galactic scales.  The proposed theory provides a new framework for addressing these anomalous electromagnetic fields.  The theory outlined above predicts electromagnetic fields when $f_{\mu\nu}\neq0$.  This condition can be the result of the usual electromagnetic source term in the action $j_{\mu}\psi^{\mu}$, where $j_{\mu}$ is the electromagnetic four-current.  However, one must also recognize the fact that the source term cannot be prescribed arbitrarily, and must be defined to be consistent with the quantity $f_0$.  Therefore, one can view the constraint as the `source' of the electromagnetic field and consequently derive the structure of astrophysical electromagnetic fields in order to satisfy the constraint in the field equations.  When $f_0$ is non-zero then $f_{\mu\nu}$ must also be non-zero.  In addition, even when $f_0$ vanishes $f_{\mu\nu}$ can be nonzero.  Therefore, either $j_{\mu}$ or $f_0$ can be considered the `source' of the electromagnetic field.  A deeper understanding of the constraint can provide insight into the origin of seemingly anomalous astrophysical electromagnetic fields.

In summary, the proposed reformulation of general relativity provides a rich framework for addressing a number of outstanding problems in contemporary gravitational physics, including: the unification of gravitation and electromagnetism; the resolution of the time discontinuity paradox; the removal of closed timelike curves from solutions of Einstein's equations; dark energy and other cosmological problems; the dark matter problem; the Pioneer anomaly; the problem of quantum gravity; and, the existence of anomalous astrophysical electromagnetic fields.  Since the proposed theory depends on both a Lagrange multiplier field $\lambda_0$ as well as the externally-defined quantity $f_0$, more work is needed to give it more predictive capabilities.  This points to the need to develop a complete theory of quantum gravity based on the proposed theory.

\bibliographystyle{unsrt}
\bibliography{qr,de,urdm,gs}

\end{document}